\documentstyle[amssymb,aps,prb]{revtex}

\begin{document}
\author{V.V.Kabanov, J. Demsar, B. Podobnik, D.Mihailovic}
\address{Jozef Stefan Institute, Jamova 39, Ljubljana, Slovenia}
\title{Quasiparticle relaxation dynamics in superconductors with different gap
structures: theory and experiments on YBa$_{2}$Cu$_{3}$O$_{7-\delta }$}
\maketitle

\begin{abstract}
Photoexcited quasiparticle relaxation dynamics are investigated in a YBa$%
_{2} $Cu$_{3}$O$_{7-\delta }$ superconductor as a function of doping $\delta 
$ and temperature $T$\ using ultrafast time-resolved optical spectroscopy. A
model calculation is presented which describes the temperature dependence of
the photoinduced quasiparticle population $n_{pe}$, photoinduced
transmission ${\it \Delta }{\cal T}/{\cal T}$ and relaxation time $\tau $
for three different superconducting gaps: $(i)$ a temperature-dependent
collective gap such that ${\bf \Delta }(T)\rightarrow 0$ as $T\rightarrow
T_{c}$, $(ii)$ a temperature-{\em independent} gap, which might arise for
the case of a superconductor with pre-formed pairs and $(iii)$ an
anisotropic (e.g. $d$-wave) gap with nodes. Comparison of the theory with
data of photoinduced transmission $\left| {\it \Delta }{\cal T}/{\cal T}%
\right| $, reflection $\left| {\it \Delta }{\cal R}/{\cal R}\right| $ and
quasiparticle recombination time $\tau $ in YBa$_{2}$Cu$_{3}$O$_{7-\delta }$
\ over a very wide range of doping ($0.1<\delta <0.48)$ is found to give
good quantitative agreement with a temperature-{\em dependent} BCS-like
isotropic gap near optimum doping ($\delta $ $<0.1)$ and a temperature-{\em %
independent }isotropic gap in underdoped YBa$_{2}$Cu$_{3}$O$_{7-\delta }$ ($%
0.15<\delta <0.48)$. A pure $d$-wave gap was found to be inconsistent with
the data.
\end{abstract}

\section{Introduction}

Ultrafast optical time resolved experiments performed on high-$T_{c}$
superconductors in recent years have shown that a substantial transient
change of the optical transmission or reflection can be induced in these
materials by ultrashort laser pulse photoexcitation (PE). What makes these
studies particularly interesting is the fact that in optimally doped
superconductors where $T_{c}$ is a maximum, the amplitude of the observed
photoinduced transmission signal, $\left| {\it \Delta }{\cal T}/{\cal T}%
\right| ,$ (or reflection $\left| {\it \Delta }{\cal R}/{\cal R}\right| 
{\cal )}$ appears to increase dramatically below $T_{c}$\cite
{Han,Chwalek,Albrecht,Stevens}. Furthermore, recently it was found that in
underdoped YBa$_{2}$Cu$_{3}$O$_{7-\delta }$ ($\delta >0.15)$, the increase
in amplitude could not be correlated with $T_{c}$, but rather with the
so-called ''pseudogap'' temperature $T^{*}$ which increases with increasing $%
\delta $\cite{Demsar}. These observations together suggest rather strongly
that the photoinduced effects are related to the opening of a gap (or
pseudogap) in the density of states close to the Fermi energy. For optimally
doped cuprate superconductors - on which most experiments were done so far -
there has been substantial agreement in the literature regarding the
experimental data. However, various groups have proposed very different
explanations for the observed effects\cite{Han,Albrecht,Stevens,Mazin}and
their origin have remained controversial\cite{comment}.

In this paper we present a calculation of the photoinduced optical response
in a superconductor for weak excitation and compare our results with
systematic experimental data on YBa$_{2}$Cu$_{3}$O$_{7-\delta }$ as a
function of $\delta $ and $T$. We derive expressions for the temperature
dependence of the photoexcited quasiparticle density $n_{pe}$ and
photoinduced transmission amplitude $\left| {\it \Delta }{\cal T}/{\cal T}%
\right| $\ for different possible gaps (or pseudogaps) which might be
applicable to high-$T_{c}$ cuprates. The simplest case considered is that of
a gap, whose temperature dependence is mean-field like and is approximated
using a BCS function. The second case considered is for a
temperature-independent ''pseudogap'' which is relevant for a superconductor
where condensation of pre-formed pairs into a superconducting state occurs
at $T_{c}$ but pairing takes place above this temperature. In this case the
gap is not{\em \ }due to a collective effect, but essentially represents the
pair binding energy $E_{B}$ and is - to first approximation - temperature
independent. Finally we discuss the case of an anisotropic gap with nodes,
such as one might expect for a $d$-wave superconductor. In section V we
present a calculation of the temperature dependence of the quasiparticle
relaxation rate $\tau $. Comparisons of the theoretical predictions with
experimental data are made in each case and a detailed discussion is given
in section VI.

\section{Experimental details}

The time-resolved experiments were performed using 100 fs, 800 nm pulses
from a Ti:sapphire laser using single high frequency modulation of the pump
at 200 kHz. The YBa$_{2}$Cu$_{3}$O$_{7-\delta }$ thin film samples were
grown on MgO or SrTiO$_{3}$ substrates and annealed in oxygen to obtain
different O concentrations $\delta .$ The $T_{c}$ was measured by measuring
the AC susceptibility in each case. The transition widths, defined as the
temperature where $\chi $ drops to 90\% of full diamagnetism, were typically
1-2 K for the $\delta <$ 0.15 and 4-7 K for $\delta >$ 0.15. The oxygen
concentration was determined from $T_{c}$, which brings in some uncertainty
near optimum doping, since the optimum $T_{c}$ for YBCO is 90 K on SrTiO$%
_{3} $ substrates and 89 K on MgO substrates, which is 2-3 K lower than in
single crystals. The typical film thicknesses were 100-120 nm and all
experiments were performed in transmission through the sample. The typical
energy of the laser pulse is 0.2 nJ, and the laser spot size typically 100$%
\mu $m in diameter. In estimating the carrier density, the absorption length
was assumed to be ${\it l}\simeq 0.1$ $\mu $m. The CW\ temperature rise of
the superconductor film due to laser heating was calibrated using a 30$\mu m$
YBCO superconducting microbridge from the same batch and on the same
substrate as in the optical experiments by measuring the change in
resistivity with a 4-probe contact method. This temperature offset was then
taken into account in the plots shown in Figures 1-6. The error in the
sample temperature is thus reduced to less than $\pm $2 K.

The photoinduced transmission ${\it \Delta }{\cal T}/{\cal T}$ \ through two
different films with $T_{c}=90$ and 53 K respectively is shown in Fig. 1 at
different temperatures above and below $T_{c}$. There are two decay
components in each case, a fast component with $\tau \sim 0.3-2$ps, and a
longer-lived component with $\tau _{L}>10$ ns, which appears as a nearly
flat background already discussed elsewhere\cite{Stevens}. The latter will
not be analysed here, except that it will be taken into account in the
fitting procedure. The time-evolution of the traces was fitted (shown by the
solid lines in Fig.1) using a model with a {\em single} exponential decay
and a Gaussian temporal profile pump pulse, from which the amplitude of the
photoinduced transmission, $\left| {\it \Delta }{\cal T}/{\cal T}\right| ,$\
and relaxation time of the fast component, $\tau ,$ was determined. The
temperature dependence of $\left| {\it \Delta }{\cal T}/{\cal T}\right| $\
and $\tau $ derived from the fits are plotted in Figs. 2, 4 and 5 and will
be discussed together with the theory in the following sections.

\section{Initial photoexcited carrier relaxation}

The initial phase of PE\ carrier relaxation after absorption of a pump laser
photon proceeds very rapidly. After the laser pulse excites an electron-hole
pair, these PE\ carriers thermalize among themselves with a characteristic
time of $\tau _{e-e}\sim \frac{\hbar E_{F}}{2\pi E^{2}}$, where $E$ is the
carrier energy measured from\ the Fermi energy $E_{F}$. In this
thermalization process, quasiparticle avalanche multiplication due to
electron-electron collisions takes place as long as $\tau _{e-e}$ is less
than the electron-phonon ({\em e-ph}) relaxation time $\tau _{e-ph}$.
Electron-phonon relaxation becomes important when the quasiparticle (QP)
energy is reduced to $E=\sqrt{\frac{\hbar E_{F}}{2\pi \tau _{e-ph}}}\simeq
30-50$meV (assuming $E_{F}\simeq 0.1-0.2$eV). $\tau _{e-ph}$ has been
determined experimentally for the case of a YBa$_{2}$Cu$_{3}$O$_{7-\delta }$%
\cite{Chekalin} as well as Bi$_{2}$Sr$_{2}$CaCu$_{2}$O$_{8+x}$ and Bi$_{2}$Sr%
$_{2}$Ca$_{2}$Cu$_{3}$O$_{10+x}$\cite{Brorson} from relaxation time fits in
time resolved experiments using intense laser pulses. Using Allen's formula 
\cite{Allen} the {\em e-ph} relaxation time for YBCO has been found to be $%
\tau _{e-ph}$ $=\frac{T_{e}}{3\lambda \left\langle \omega ^{2}\right\rangle }%
\simeq 100$fs for initial carrier temperatures $T_{e}=E_{I}/C_{e}$ in the
range $3000$ K\cite{Chekalin} and $\tau _{e-ph}$ $\simeq 60$fs for $%
T_{e}\simeq 410$ K \cite{Brorson}. Here $E_{I}$ is the energy density per
unit volume deposited by the laser pulse, $C_{e}$ is the electronic specific
heat, $\lambda $ is a constant which characterizes the electron-phonon
interaction and $\left\langle \omega ^{2}\right\rangle $ is a mean-square
phonon frequency. $\lambda $ has been determined from these experiments to
be in the range $0.9<\lambda <1$ both in YBa$_{2}$Cu$_{3}$O$_{7-\delta }$
and Bi$_{2}$Sr$_{2}$CaCu$_{2}$O$_{8+x}$ superconductors. In experiments
which we are considering here\cite{Han,Stevens,Demsar}, the laser
intensities are significantly smaller, and the photoexcited electronic
temperature $T_{e}$ is only few K in excess of the lattice temperature, i.e. 
$T_{e}$ $\simeq T,$ \ which gives an {\em e-ph} relaxation time $\tau
_{e-ph}\simeq 10$fs. Under these near-equilibrium conditions where the
carrier temperature is near the lattice temperature, this value of $\tau
_{e-ph}\ $can be compared to the relaxation time determined from infrared
reflectivity measurements. For energies of the order of $30-50$ meV, $\hbar
/\tau $ $\sim $ 3000 cm $^{-1}$ ($\tau \sim 16$ fs$)$\cite{Basov} in good
agreement with the estimated $\tau _{e-ph}\simeq 10$fs. In the absence of a
gap in the low-energy density of states, this is the timescale of the
electron -phonon thermalization.

In the presence of the gap the situation is strongly modified and a
bottleneck in the relaxation occurs after $t\geq 10$ fs. As a result QPs
accumulate near the gap, forming a non-equilibrium distribution. Each photon
thus creates $30\sim 40$ QPs given by $E_{pump}/2{\bf \Delta ,}$where $%
E_{pump}$ $=1.5$eV is the laser pump photon energy and $2{\bf \Delta }$ is
the energy gap.\ Because the final relaxation step across the gap is
strongly suppressed \cite{Roth,Aronov} the QPs together with{\em \ }high
frequency phonons (with $\hbar \omega >2{\bf \Delta })$ form a near-steady
state distribution. The QP\ recombination dynamics of this system is
governed by the emission and\ reabsorption of high frequency phonons.
Phonons with $\hbar \omega <2{\bf \Delta }$ do not participate in the direct
relaxation of the QPs, since in this case the QP\ final states would lie in
the gap. (A quantitative justification for the assumption that phonons play
a dominant role in the QP relaxation will be given in section V.) The basis
for these pump-probe experiments is that this near-equilibrium QP population
can be very effectively probed using excited state absorption using a
suitable second optical probe pulse, giving direct information on QP
dynamics and the nature of the gap itself.

Since the probe laser photon energy $E_{probe}$ is typically {\em well above}
the plasma frequency in high-$T_{c}$ superconductors, we make the
approximation that the transition probability for the probe light is given
by the Fermi golden rule, with photoinduced quasiparticles as initial states%
{\em \ }and with final unoccupied states well above the Fermi energy in a
band which lies approximately at $E_{0}\sim E_{probe}$. The amplitude of the
photoinduced absorption (PA) $\left| {\it \Delta }{\cal A}/{\cal A}\right| $
is then proportional to the photoinduced transmission $\left| {\it \Delta }%
{\cal T}/{\cal T}\right| $ (and in the linear approximation also to $\left| 
{\it \Delta }{\cal R}/{\cal R}\right| {\cal )}$ which is in turn
proportional to the number of photoexcited quasiparticles $n_{pe}.$ The
probe signal is thus weighted by the O-Cu charge-transfer dipole matrix
element and the joint DOS, so $\left| {\it \Delta }{\cal T}/{\cal T}\right|
\propto -n_{pe}\rho _{f}\left| M_{ij}\right| ^{2}$ where $n_{pe}$ is the
photoexcited carrier density, $\rho _{f}$ is the final density of unoccupied
states and $M_{ij}=\left\langle {\bf p.A}\right\rangle $ is the dipole
matrix element. Although recent experiments on YBa$_{2}$Cu$_{3}$O$_{7-\delta
}$ with different $E_{probe}$ show the existence of a resonance for $%
E_{probe}\sim 1.5$ eV\cite{Stevens}, we assume here that the adiabatic
approximation can be applied, so the photoinduced carriers do not cause any
change in $\rho _{f}$ or $M_{ij}.$ In this case the effect of the 1.5 eV
resonance is to significantly enhance the sensitivity of the probe.

\section{Temperature dependence of the quasiparticle density}

\subsection{Isotropic gap.}

In discussing the theoretical explanation for the observed effects, it is
important to make the distinction between experiments {\em (i)} in which the
photoexcited charge carrier density is substantially less than the normal
state carrier density, $n_{pe}$ $\ll n_{c}$ so the laser makes only a weak
perturbation on the superconductor\cite{Han,Stevens,Demsar} and {\em (ii)}
those where $n_{pe}\sim n_{c}$ and photoexcitation is sufficiently strong to
close the superconducting gap\cite{Chwalek,Albrecht}. Whereas for the latter
case {\em (ii)}, a theoretical description was proposed by Mazin\cite{Mazin}%
, so far there has been no theoretical calculation for the photoinduced
effects in a superconductor for the case of weak photoexcitation {\em (i)}.
The number of photogenerated quasiparticles in low-excitation density
experiments is $n_{pe}\lesssim 3\times 10^{-3}$/unit cell\cite
{Han,Stevens,Demsar}. On the other hand, the typical quasiparticle
concentration in a high-$T_{c}$ superconductor is $n_{0}=2N(0){\bf \Delta }%
\simeq 0.2-0.4$/unit cell, where $N(0)$ is the density of states at $E_{F}.$
The number of photoexcited QPs is small compared to the normal state density 
$n_{pe}/n_{0}\lesssim 10^{-2}$, so the weak photoexcitation approximation is
clearly justified and the photoexcited quasiparticles make only a small
perturbation of the distribution functions. Assuming that the energy gap is
more or less isotropic (no nodes), we can approximate nonequilibrium phonon (%
$n_{\omega _{q}})$ and quasiparticle ($f_{\epsilon })$ distribution
functions as follows \cite{Aronov1}: 
\begin{equation}
n_{\omega _{q}}= 
{\ \ \frac{1}{\exp (\frac{\hbar \omega _{q}}{k_{B}T})-1}\quad \quad \hbar \omega _{q}<2{\bf \Delta } \atop \frac{1}{\exp (\frac{\hbar \omega _{q}}{k_{B}T^{^{\prime }}})-1}\quad \quad \hbar \omega _{q}>2{\bf \Delta }}%
\end{equation}
\begin{equation}
f_{\epsilon }=\quad \frac{1}{\exp (\frac{\epsilon }{k_{B}T^{^{\prime }}})+1}
\end{equation}
where $T$ is the lattice temperature and $T^{^{\prime }}$ is the temperature
of quasiparticles and high frequency phonons with $\hbar \omega _{q}>2{\bf %
\Delta }$.

We can calculate $n_{pe}$ using Eqs.(1), (2) and by considering the
conservation of energy. Assuming ${\bf \Delta }$ is temperature {\em %
independent} and large in comparison to $T,$ the conservation of energy has
the following form: 
\begin{equation}
\left( n_{T^{^{\prime }}}-n_{T}\right) {\bf \Delta }+\left( n_{T^{^{\prime
}}}^{2}-n_{T}^{2}\right) \frac{\nu {\bf \Delta }}{2\hbar \Omega
_{c}N(0)^{2}k_{B}T^{^{\prime }}}={\cal E}_{I}
\end{equation}
here $n_{T^{^{\prime }}},n_{T}$ is the number of thermally excited
quasiparticles per unit cell at $T^{^{\prime }}$ and $T$ respectively, $%
{\cal E}_{I}$ is the energy density per unit cell deposited by the incident
laser pulse, $\Omega _{c}$ is phonon frequency cut off and $\nu $ is the
effective number of phonon modes per unit cell participating in the
relaxation. Taking into account that $n_{pe}=(n_{T^{^{\prime }}}-n_{T})\ll
n_{T}$ and assuming that $n_{T}=2N(0)k_{B}T\exp (-{\bf \Delta }/k_{B}T)$,
the number of photogenerated quasiparticles at temperature $T$ is given by:

\begin{equation}
n_{pe}=\frac{{\cal E}_I/{\bf \Delta }}{1+\frac{2\nu }{N(0)\hbar \Omega _c}%
\exp (-{\bf \Delta }/k_BT)}.
\end{equation}
For the case of a temperature-dependent gap ${\bf \Delta }(T)$, Eq.(3) will
be slightly modified:

\begin{equation}
\left( \left( n_{T^{^{\prime }}}-n_{T}\right) +\frac{\nu }{\hbar \Omega
_{c}N(0)^{2}\pi {\bf \Delta }(T)}\left( n_{T^{^{\prime
}}}^{2}-n_{T}^{2}\right) \right) ({\bf \Delta }(T)+k_{B}T/2)={\cal E}_{I}
\end{equation}
We approximate $n_{T}\simeq 2N(0)\sqrt{\pi {\bf \Delta }(T)k_{B}T/2}\exp (-%
{\bf \Delta }(T)/T)$ \cite{Aronov} with $k_{B}T\ll {\bf \Delta }$, resulting
in a slightly modified expression for $n_{pe}$\cite{approx}:

\begin{equation}
n_{pe}=\frac{{\cal E}_{I}/({\bf \Delta }(T)+k_{B}T/2)}{1+\frac{2\nu }{%
N(0)\hbar \Omega _{c}}\sqrt{\frac{2k_{B}T}{\pi {\bf \Delta }(T)}}\exp (-{\bf %
\Delta }(T)/k_{B}T)}.
\end{equation}

We note that in Eqs. (4) and (6), the explicit form of $n_{pe}\,($ and
hence\ also $\left| {\it \Delta }{\cal A}/{\cal A}\right| $ or $\left| {\it %
\Delta }{\cal T}/{\cal T}\right| {\cal )}$ depends only on the ratio $k_{B}T/%
{\bf \Delta }$, showing that the intensity of the photoresponse is a
universal function of $k_{B}T/{\bf \Delta }$ as long as the particular
functional form of the temperature dependence ${\bf \Delta }(T)$ is the
same. Another important feature of the expressions (4) and (6) is that at $%
T=0$, $n_{pe}\propto 1/{\bf \Delta (}0{\bf )}$, which allows a determination
of ${\bf \Delta }(0)$, as long as the experimental parameters required to
calculate ${\cal E}_{I}$ are recorded sufficiently precisely.

Since the decay of each phonon involves the creation of two quasiparticles,
the number of high frequency phonons is proportional to the square of the
number of quasiparticles $n_{T^{^{\prime }}}^{2}$ at temperature $%
T^{^{\prime }}$. (Taking into account conservation of energy\cite{Aronov1},
at higher pulse energies when $n_{pe}$ becomes large, we expect a crossover
of the photoresponse from a linear dependence on laser power, $n_{pe}\propto
I$ to a square root dependence $n_{pe}\propto \sqrt{I}$ at very high
intensities.)

To enable comparison of the temperature dependences of the photoinduced
transmission given by Eqs. (4) and (6) with experiments, we have plotted
them as a function of temperature in Figures 2b) and c) respectively. The
values for the constants for YBa$_{2}$Cu$_{3}$O$_{7-\delta }$ are: $\nu
=10-20$, $N(0)=2.5-5$ eV$^{-1}$spin$^{-1}$cell$^{-1}$, $\Omega _{c}=0.1$ eV.
At low temperatures $n_{pe}$ is essentially $T$-independent in both cases,
falling off at higher temperatures$.$ However, from the figure it is clear
that the high-temperature behaviour is quite different in the two cases. In
the case of the $T$-dependent gap (Fig.2c)), the decrease in $\left| {\it %
\Delta }{\cal T}/{\cal T}\right| $ is much more pronounced and starts quite
close to $T_{c}$ at $T/T_{c}\approx 0.8$ dropping to zero {\em at} $T_{c}.$
In the temperature-independent gap case (Fig. 2b)), the photoinduced
transmission starts to drop at much lower temperatures near $T/T^{\ast
}\approx 0.4,$ dropping exponentially at high temperatures. Also, a notable
prediction of Eq. (6) for the $T$-dependent gap is a small peak near $%
T/T_{c}\approx 0.7,$ which is not present in the case of a
temperature-independent gap formula (4).

In Fig. 2a) we have plotted the data for the photoinduced transmission
amplitude for YBa$_{2}$Cu$_{3}$O$_{7-\delta }$ as a function of temperature
for three different $\delta .$ Both the magnitude and the temperature
dependence are seen to be strongly doping-dependent. For comparison with
theory, superimposed on the theoretical curves in Figure 2 b) and c) we have
plotted the normalised photoinduced reflectivity and transmission data for
YBa$_{2}$Cu$_{3}$O$_{7-\delta }$ with different $\delta $. The data for a
number of underdoped samples\cite{Demsar} in the range $0.15<\delta <0.48$
have been scaled onto a common temperature scale as suggested by the formula
(4) and plotted as a function of $T/T^{\ast }$ in Figure 2b). $T^{\ast }$ is
defined as the point where the amplitude of\ signal $\left| {\it \Delta }%
{\cal T}/{\cal T}\right| $ drops to some fixed value (for example to 5\% of
the maximum value). The data can be seen to fit the theoretical curve
remarkably well.

For near-optimally doped YBa$_{2}$Cu$_{3}$O$_{7-\delta }$ $(\delta <0.1)$
the data are scaled by $T_{c}$ and are plotted as a function of reduced
temperature $T/T_{c}$. However, this time the data are superimposed on the
prediction for the $T$-dependent BCS-like gap Eq. (6) in Figure 2c). In this
case also, the fit is seen to be good, including the small maximum at $%
T/T_{c}\approx 0.7$. This maximum is particulary well observed in the data
by Han et al.\cite{Han} and by Mihailovic et al.\cite{Demsar}, although it
is not evident in the data of Stevens et al.\cite{Stevens}.

The experimental data thus appear to scale onto {\em one} of the {\em two}
theoretical curves, the near-optimally doped YBa$_{2}$Cu$_{3}$O$_{7-\delta }$
data $(\delta <0.1)$ agreeing very well with Eq. (6) derived for a
temperature-dependent gap, while the underdoped data ($\delta >0.15)$ fit
expression (4) for a $T$-independent gap. The fact that the underdoped
sample data scale onto one universal curve in Figure 2b), while the
near-optimally doped data scale onto the curve in Fig. 2c), is a consequence
of the scaling properties of Eqs.(4) and (6) and is a rather remarkable
confirmation of the theoretical model by the experiment. From the fits of
the underdoped sample data, we obtain gap values of $2\Delta =(5\pm
1)k_{B}T^{\ast }$ for underdoped samples and $2\Delta =(9\pm 1)k_{B}T_{c}$
for the near-optimally doped samples, the latter in good agreement with
other optical experiments\cite{Basov,Schles}.

Further confirmation of the model comes from the predicted linear intensity
dependence in Eqs (4) and (6). In Figure 3 we plot the $T$-dependence of $%
\left| {\it \Delta }{\cal T}/{\cal T}\right| $ for near optimally doped
sample ($T_{c}$ = 89K) for three different pump intensities. The data are
seen to scale linearly with intensity as shown by the insert where ${\cal E}%
_{I}$ is changed over one order of magnitude. (Unfortunately, measurements
over a larger range of intensities were not possible because of steady-state
laser heating.)

\subsection{Anisotropic gap with nodes}

If the gap is anisotropic and contains nodes on the Fermi surface - such as
for $d$-wave or strongly anisotropic $s$-wave pairing - there is no clear
gap in the spectrum. Nevertheless the quasiparticle DOS still has a strong
energy dependence and a separation of low and high energy quasiparticles
still exists, albeit is less pronounced than in the isotropic case. For an
anisotropic gap with nodes we can approximate the quasiparticle DOS as a
function of energy $\varepsilon $ as: 
\begin{equation}
N(\varepsilon )=N(0)\left( \frac{\varepsilon }{{\bf \Delta }_{a}}\right)
^{\eta }\ \ (\varepsilon \ll {\bf \Delta }_{a}),
\end{equation}
where ${\bf \Delta }_{a}$ is a characteristic energy scale separating the
low and high-energy QPs, and the exponent $\eta $ depends on the topology of
the nodes on the Fermi surface. For a two dimensional Fermi surface (with
nodes) $\eta =1$, while for a 3-dimensional case $\eta =2$.

Using equation (7) we can estimate the average energy accumulated by
quasiparticles after each laser pulse: 
\begin{equation}
{\it \Delta }E=2FN(0)(k_{B}T^{\prime (\eta +2)}-k_{B}T^{(\eta +2)})/{\bf %
\Delta }_{a}^{\eta }={\cal E}_{I}
\end{equation}
\begin{equation}
F=\int_{0}^{\infty }\frac{x^{\eta +1}}{\exp (x)+1}dx
\end{equation}
$F=1.80$ for $\eta =1,$ and $F=5.68$ for $\eta =2$, and $T^{^{\prime }}$ and 
$T$ are the QP temperature and lattice temperatures respectively. At low
temperatures, the phonon contribution to the total energy is exponentially
small (Eq.(8)) and can be neglected. In this case, the high and low energy
phonons are separated by an energy scale of the order of ${\bf \Delta }_{a}$%
. (Note that in general ${\bf \Delta }_{a}$ scales with the gap, but is not
equal to the amplitude of the gap.)

Solving Eq.(8) with respect to the QP temperature $T^{\prime }$ we obtain: 
\begin{equation}
T^{\prime }=\left[ \frac{{\cal E}_{I}{\bf \Delta }_{a}}{2FN(0)}%
+k_{B}T^{(\eta +2)}\right] ^{1/(\eta +2)}.
\end{equation}
The number of thermally excited quasiparticles is determined by the
equation: 
\begin{equation}
n_{T}=2GN(0)k_{B}T^{(\eta +1)}/{\bf \Delta }_{a}^{\eta },
\end{equation}
where $G$ is given by the Eq.(9), except that $\eta +1$ is replaced by $\eta
,$ so $G=0.82$ for $\eta =1$ and $G=1.80$ for $\eta =2$. Combining the two
equations (10) and (11), we derive the following equation for the number of
photogenerated quasiparticles as a function of temperature: 
\begin{equation}
n_{pe}=\frac{2GN(0)k_{B}T^{(\eta +1)}}{{\bf \Delta }_{a}^{\eta }}\left[
\left( 1+\frac{{\cal E}_{I}{\bf \Delta }_{a}^{\eta }}{2FN(0)k_{B}T^{(\eta
+2)}}\right) ^{\frac{\eta +1}{\eta +2}}-1\right] ,
\end{equation}
The expression for the photoinduced carrier density Eq.(12) is plotted in
Figure 4 with $\eta =2$ (solid line) and the same values of $N(0)$ and $%
{\cal E}_{I}$ as before. Instead of an exponential fall-off with increasing
temperature, the photoexcited quasiparticle number decreases rapidly well
below $T_{c}$ according to a power law. In the 2-dimensional case with $\eta
=1,$ (and the same parameters), the temperature dependence is shown by the
dashed line in Figure 4 and differs only at the lowest temperatures. Using a 
$T$-dependent gap (e.g. of BCS form) in Eq. (12), the curve for $n_{pe}$ is
virtually indistinguishable from the case where ${\bf \Delta }_{a}^{\eta }$\
is temperature-independent. The reason for this is that the effect of the $T$%
-dependent gap is only important as $T\rightarrow T_{c},$ but there $n_{pe}$
is already small.

At low temperatures we can neglect the 1 in the round brackets and obtain: 
\begin{equation}
n_{pe}=\frac{2GN(0)}{{\bf \Delta }_{a}^{\eta }}\left[ \left( \frac{{\cal E}%
_{I}{\bf \Delta }_{a}^{\eta }}{2FN(0)}\right) ^{\frac{\eta +1}{\eta +2}%
}-k_{B}T^{(\eta +1)}\right] .
\end{equation}
A feature of the $d$-wave model - which is particularly important for
comparison with experiment - is the peculiar sub-linear behaviour of $n_{pe}$
with laser pump intensity, $n_{pe}\sim {\cal E}_{I}^{\frac{\eta +1}{\eta +2}%
} $ at low temperatures. In the high temperature limit on the other hand we
can expand the term in brackets to obtain a crossover to linear behaviour,
with a temperature-dependent slope: 
\begin{equation}
n_{pe}=2G\frac{{\cal E}_{I}}{k_{B}T}\frac{\eta +1}{\eta +2}
\end{equation}

It should be pointed out that in the high temperature limit, the phonon
contribution should become increasingly important and the decrease of the
number of photogenerated QP will eventually become exponential, just as in
the case of an isotropic gap.

The data for the temperature dependence of the induced transmission for
underdoped and optimally doped YBa$_{2}$Cu$_{3}$O$_{7-\delta }$ are also
plotted on Figure 4, superimposed with the predicted response for the $d$%
-wave case. Clearly the $T$-dependence data cannot be described using a {\em %
pure} $d$-wave gap irrespective of dimensionality or the form of temperature
dependence of ${\bf \Delta }(T)$ (Eq.(12)). The linear intensity dependence
shown in Fig. 3 is consistent with this observation, and is also not
consistent with a $d$-wave gap, which predicts a sub-linear response. We
note that although the data apparently exclude the possibility of a {\em %
pure }$d$-wave gap in underdoped and near-optimally doped YBCO, it does not
entirely rule out a mixed symmetry gap with some $d$-wave admixture.

\section{The quasiparticle relaxation rates.}

As discussed in the section III, the relaxation rate of the photoinduced QPs
is dominated by the energy transfer from high-frequency phonons with $\hbar
\omega >2{\bf \Delta }$ to phonons with $\hbar \omega <2{\bf \Delta .}$ To
describe the relaxation of nonequilibrium quasiparticles in this case we
consider the kinetic equation for phonons, taking into account phonon-phonon
scattering\cite{Lifshitz}:

\begin{equation}
\frac{\partial n_{\omega }}{\partial t}=I_{ph-ph}\{n_{\omega }\}
\end{equation}
where phonon-phonon scattering integral has the form:

\begin{eqnarray}
I_{ph-ph}\{n_{\omega }\} &=&2\pi \sum_{q_{1},q_{2}}\left|
w_{q_{1},q_{2}}\right| ^{2}\{\frac{1}{2}\left[ \left( n_{\omega }+1\right)
n_{\omega _{1}}n_{\omega _{2}}-n_{\omega }\left( n_{\omega _{1}}+1\right)
\left( n_{\omega _{2}}+1\right) \right] \delta \left( \omega -\omega
_{1}-\omega _{2}\right)  \nonumber \\
&&+\left[ \left( n_{\omega }+1\right) \left( n_{\omega _{1}}+1\right)
n_{\omega _{2}}-n_{\omega }n_{\omega _{1}}\left( n_{\omega _{2}}+1\right)
\right] \delta \left( \omega _{2}-\omega -\omega _{1}\right) \},
\end{eqnarray}
$w_{q_{1},q_{2}}$ is the anharmonic coupling constant and $n_{\omega }$ is
the phonon distribution function. (Conservation of the phonon momentum is
not relevant because of umklapp scattering.) We neglect the {\em %
electron-phonon} scattering integral, because at normal temperatures the
relative contribution of quasiparticles to the relaxation of phonons is
small due to the relatively small number of quasiparticles compared to the
phonons $N(0)\Omega _{c}\ll \nu .$ Electron-phonon collisions can play an
important role only at temperatures $T^{^{\prime }}\ll T_{c}$ but since - as
shown in the previous section - $n_{pe}$ is nearly constant at low
temperature, $T^{^{\prime }}$ cannot be small. The distribution function
Eq.(1) reduces to $0$ the parts of the collision integral Eq.(16) which
describe the scattering of high and low frequency phonons separately. Only
the part of the collision integral that does not conserve the number of
phonons with $\omega <2{\bf \Delta }$ differs from zero.

To estimate the relaxation time, we multiply Eq. (15) by $\omega _{q}$ and
sum over $q$, satisfying the condition $0<\omega _{q}<2{\bf \Delta }(T)$:

\begin{equation}
\frac{\partial E_{<}}{\partial t}=\sum_{q(\omega _{q}<2{\bf \Delta })}\hbar
\omega _{q}I_{ph-ph}\{n_{\omega }\}.
\end{equation}

If we suppose that the coupling constant is momentum independent $%
w_{q_{1},q_{2}}=w$, then:

\begin{eqnarray}
\sum_{q(\omega _{q}<2{\bf \Delta })}\omega _{q}I_{ph-ph}\{n_{\omega }\}
&=&2\pi w^{2}\int_{0}^{2{\bf \Delta }}\omega \rho \left( \omega \right)
d\omega \int \{\frac{1}{2}  \nonumber \\
&&\left[ \left( n_{\omega }+1\right) n_{\omega ^{^{\prime }}}n_{\omega
-\omega ^{^{\prime }}}-n_{\omega }\left( n_{\omega ^{\prime }}+1\right)
\left( n_{\omega -\omega ^{\prime }}+1\right) \right] \rho \left( \omega
^{^{\prime }}\right) \rho \left( \omega -\omega ^{^{\prime }}\right) + 
\nonumber \\
&&\left[ \left( n_{\omega }+1\right) \left( n_{\omega ^{^{\prime
}}}+1\right) n_{\omega +\omega ^{^{\prime }}}-n_{\omega }n_{\omega
^{^{\prime }}}\left( n_{\omega +\omega ^{^{\prime }}}+1\right) \right] \rho
(\omega ^{^{\prime }})\rho (\omega ^{^{\prime }}+\omega )\}d\omega
^{^{\prime }}
\end{eqnarray}
here $\rho \left( \omega \right) =3\nu \omega ^{2}/\Omega _{c}^{3}$ is the
phonon density of states in the Debye approximation. We restrict ourselves
to a discussion of the relatively high temperature limit where the phonon
distribution function can be replaced by $k_{B}T/\hbar \omega $. The energy
accumulated by the phonons with $\hbar \omega <2{\bf \Delta }$ is:

\begin{equation}
E_{<}=\int_{0}^{2{\bf \Delta }}n_{\omega }\rho \left( \omega \right) \hbar
\omega d\omega =\frac{8\nu k_{B}T{\bf \Delta }^{3}}{(\hbar \Omega _{c})^{3}}
\end{equation}

\smallskip In the integral Eq.(18) the first term is equal to $0,$ because
it has only phonon distribution functions with $\hbar \omega <2{\bf \Delta }$%
. The first non-zero contribution describes the decay of the phonon with $%
\hbar \omega >2{\bf \Delta }$ to two phonons with $\hbar \omega <2{\bf %
\Delta }$:

\begin{equation}
I_{1}\simeq \frac{13}{12}\pi w^{2}\left( \frac{3\nu }{(\hbar \Omega _{c})^{3}%
}\right) ^{3}k_{B}^{2}T(T^{^{\prime }}-T)(2{\bf \Delta })^{7}
\end{equation}

The second contribution describes the inelastic scattering of high frequency
phonon with creation of one low frequency phonon with $\omega <2{\bf \Delta }
$:

\begin{equation}
I_{2}\simeq \frac{8}{3}\pi w^{2}\left( \frac{3\nu }{(\hbar \Omega _{c})^{2}}%
\right) ^{3}k_{B}^{2}T^{^{\prime }}(T^{^{\prime }}-T){\bf \Delta }^{4}
\end{equation}

Eq.(21) shows that the inelastic scattering of high frequency phonons is
dominant in the energy relaxation of the energy if ${\bf \Delta }\ll \hbar
\Omega _{c}$ and we can neglect $I_{1}$.

Taking into account Eqs.(17), (19-21) we obtain expressions describing the
relaxation of the equilibrated QP-phonon temperature: 
\begin{equation}
\frac{\partial T}{\partial t}=\frac{1}{\tau _{ph}}(T^{^{\prime }}-T)
\end{equation}
\begin{equation}
\frac{1}{\tau _{ph}}=\frac{9\pi \nu ^{2}w^{2}k_{B}T^{^{\prime }}{\bf \Delta }%
(T)}{(\hbar \Omega _{c})^{3}}
\end{equation}
Note that Eq.(22) is rather similar to that derived by Allen\cite{Allen} for
the temperature relaxation in the electron-phonon system in the normal
metals, except that Eq.(22) describes the phonon energy relaxation modified
by the gap in the quasiparticle spectrum.

The phonon-phonon relaxation time (Eq.23) can be expressed in terms of an
experimental parameter, namely the Raman phonon linewidth $\Gamma _{\omega
}. $ Using the Fermi golden rule we can calculate $\Gamma _{\omega }$ for $%
k_{B}T\ll \hbar \omega $: 
\begin{equation}
\Gamma _{\omega }=2\pi w^{2}\sum_{q,\nu ,\nu ^{^{\prime }}}\left( n_{\omega
_{q,\nu }}+1\right) \left( n_{\omega _{-q,\nu ^{^{\prime }}}}+1\right)
\delta (\omega _{q,\nu }+\omega _{-q,\nu ^{^{\prime }}}-\omega )\simeq \frac{%
3\pi w^{2}\nu ^{2}\omega ^{2}}{4\hbar \Omega _{c}^{3}}.
\end{equation}

The phonon relaxation rate can thus be expressed in terms of $\Gamma
_{\omega }$ in the following form: 
\begin{equation}
\frac{1}{\tau _{ph}}=\frac{12\Gamma _{\omega }k_{B}T^{^{\prime }}{\bf \Delta 
}(T)}{\hbar \omega ^{2}}
\end{equation}

For a temperature dependent gap ${\bf \Delta },$ the relaxation time is
expected to show a divergence $\tau _{ph}\propto 1/{\bf \Delta }(T)$ as is
indeed observed in optimally doped cuprates\cite{Han,Demsar,Brorson,Easley}.
A similar divergence of the relaxation time has been calculated previously
by Schmidt and Sch\"{o}n\cite{Schmidt} and Tinkham\cite{Tinkham}, albeit for
somewhat differently created nonequilibrium situations.

The typical relaxation timescale $\tau _{ph}$ given by the formula Eq.(25)
is very close to the experimentally observed values. From the data on the
Raman linewidth of the $A_{1g}$-symmetry apical O(4) phonon mode in YBCO -
which has been shown to be particularly anharmonic - $\Gamma _{\omega
}\approx 13$ cm$^{-1}$ and $\omega \simeq 400\ $cm$^{-1}$\cite{Mihailovic1}.
At $T^{^{\prime }}\simeq T=T_{c}/2$, and using ${\bf \Delta }%
_{T_{c}/2}\approx 200$ cm$^{-1}$ we obtain $\tau _{ph}=$ 0.8 ps.

At low temperatures the quasiparticle temperature $T^{^{\prime }}$ is much
higher than the lattice temperature $T$ and the formula is expected to fail.
However, from the experiments (Figure 2) we see from the temperature
dependence of $\left| {\it \Delta }{\cal T}/{\cal T}\right| $\ that the
number of photogenerated quasiparticles is nearly constant at low $T$. $%
T^{^{\prime }}$ can therefore be estimated from the equation: 
\begin{equation}
n_{T^{^{\prime }}}-n_{T}={\cal E}_{I}/{\bf \Delta }(0)
\end{equation}
Taking into account that $n_{T}\simeq 2N(0){\bf \Delta }(0)\exp (-{\bf %
\Delta }(T)/k_{B}T)$ we obtain: 
\begin{equation}
k_{B}T^{^{\prime }}\simeq {\bf \Delta }(T)/\ln \left\{ 1/({\cal E}_{I}/2N(0)%
{\bf \Delta }(0)^{2}+\exp (-{\bf \Delta }(T)/k_{B}T))\right\}
\end{equation}

At low $T,$ the nonequilibrium temperature $T^{^{\prime }}\simeq {\bf \Delta 
}(T)/\ln \left\{ (2N(0){\bf \Delta }(0)^{2}/{\cal E}_{I}\right\} \simeq
T_{c}/2$, while in the high temperature limit, the exponent in the logarithm
becomes large and $T^{^{\prime }}\simeq T$. Combining Eq.(25) and Eq.(27) we
obtain an expression for the QP\ relaxation time as a function of lattice
temperature $T$ and photoexcitation energy ${\cal E}_{I}$ which is valid for
all temperatures $0<T<T_{c}$:

\begin{equation}
\frac{1}{\tau _{ph}}=\frac{12\Gamma _{\omega }{\bf \Delta }(T)^{2}}{\hbar
\omega ^{2}\ln \left\{ 1/({\cal E}_{I}/2N(0){\bf \Delta }(0)^{2}+\exp (-{\bf %
\Delta }(T)/k_{B}T))\right\} }.
\end{equation}

To enable comparison of this formula with experiments$,$ we again use Raman
data on high-frequency optical phonon linewidths, a value of gap frequency
of 200 cm$^{-1}$ $(2{\bf \Delta }\simeq 9k_{B}T_{c})$ from the fits in
Figure 2c), the same values of $N(0)=2.2-5$ eV$^{-1}$cell$^{-1}$spin$^{-1}$
as in section III, and a BCS function for ${\bf \Delta (}T{\bf )}$. ${\cal E}%
_{I}$ was again calculated using an incident laser energy of 0.2 nJ incident
on a 100 nm thick film with a spot size of $\sim 100\mu $m diameter. We
obtain a temperature dependence of the relaxation time $\tau _{ph}$ as shown
in Figure 5. A comparison of the theory with the data for optimally doped YBa%
$_{2}$Cu$_{3}$O$_{7-\delta }$ samples\cite{Han,Demsar} in Figure 5 shows
good agreement with the theory\cite{divergence} with {\em no adjustable
parameters}, especially near $T_{c}$. Some experimental data in the
literature\cite{Han,Reitze} show an upturn in relaxation time at low
temperatures, which is also weakly present in the data in Fig. 5. Although
Eq.(28) is derived for temperatures close to $T_{c},$ an upturn at low
temperature can still be reproduced by this formula if we reduce the energy
per pulse ${\cal E}_{I}$. (In fact in the limit of ${\cal E}_{I}\rightarrow
0,$ Eq. (28) gives $\tau _{ph}$ $\propto $ $1/T$ at low $T$.) Indeed by
setting ${\cal E}_{I}/N(0)$ in Eq. (28) as an adjustable parameter, rather
than inserting fixed values as we have done, the low-temperature upturn in
the data can be reproduced more accurately.

To end this section we briefly make a quantitative comparison of the
relaxation rates for a gapped and gapless material. Comparing the relaxation
time for the case of a gapped material with relaxation time in a normal
metal described by Allen's formula\cite{Allen} we obtain: 
\begin{equation}
\frac{\tau _{ph}}{\tau _{e-ph}}\sim \frac{\lambda \hbar ^{3}\omega ^{4}}{%
4\pi (k_{B}T)^{2}{\bf \Delta }\Gamma _{\omega }}
\end{equation}
Inserting experimental values for $\omega $ and $\Gamma _{\omega }$ we find
the anharmonic phonon decay time in a superconductor to be approximately two
orders of magnitude {\em slower }than electron-phonon relaxation in a
gapless material as $\tau _{ph}/\tau _{e-ph}\sim 60-200$ for $T\simeq T_{c}$
as expected.

The calculation was performed assuming that the rate limiting step for QP
relaxation is anharmonic decay of high-energy phonons, while direct electron
phonon relaxation was assumed to be small. This is now justified
quantitatively by a calculation of the direct electron-phonon relaxation
rate. The contribution of phonon-electron scattering to the relaxation of
phonon system can be estimated by deriving the corresponding relaxation rate 
$\tau _{1}^{-1}$ using the usual electron-phonon collision integral (see
Appendix): 
\begin{equation}
\frac{1}{\tau _{1}}=\frac{\pi N(0)\lambda {\bf \Delta }(T)\Omega _{c}}{2\nu }
\end{equation}
We see that the electron-phonon relaxation rate, when compared to the phonon
relaxation rate is reduced by the factor $N(0)\hbar \Omega _{c}/\nu \ll 1$: 
\begin{equation}
\frac{\tau _{ph}}{\tau _{1}}=\frac{\pi \hbar ^{2}N(0)\lambda \omega
^{2}\Omega _{c}}{24\nu \Gamma _{\omega }k_{B}T^{^{\prime }}}<1
\end{equation}
and can therefore be neglected. However if the number of phonon modes
involved in the relaxation is significantly smaller, the ratio $\tau
_{ph}/\tau _{1}$ may approach unity. In this case the electron-phonon
collisions can also contribute to the phonon relaxation. However the
temperature dependence of the relaxation rate $1/\tau _{1}$ is very similar
to the anharmonic relaxation $\tau _{ph}$, both being proportional to ${\bf %
\Delta }(T$) and so the two contributions to the relaxation may be difficult
to separate experimentally. The data shown in Figure 5 are fit to $\tau
_{ph} $ without any fitting parameters, so it appears to be reasonable to
assume that anharmonic relaxation\ described by Eqs. (25) and (28) is
dominant in YBa$_{2}$Cu$_{3}$O$_{7-\delta }$ near optimum doping.

\section{Discussion}

Near optimum doping in YBa$_{2}$Cu$_{3}$O$_{7-\delta }$ $(\delta <0.1)$ we
find that the theoretical model gives a good quantitative fit to the
temperature dependence of the photoinduced transmission (Figure 2c)),
provided we use a temperature-dependent gap (Eq. (6)). The relatively square
shape of the temperature dependence of $\left| {\it \Delta }{\cal T}/{\cal T}%
\right| $ , the weak maximum near $T/T_{c}=0.7$ and the rapid drop in the
signal amplitude above this temperature is a consequence of the particular
temperature dependence of the gap with the property that ${\bf \Delta (}T%
{\bf )}\rightarrow 0$ as $T\rightarrow T_{c}$. In underdoped YBCO on the
other hand (Fig. 2b)), the temperature dependence is much weaker and can
only be reproduced by the calculation using a temperature-independent gap
(Eq. (4)). It is {\em not possible }to reproduce the temperature dependence
of the optimally doped data using a temperature-independent gap and vice
versa, it is not possible to describe the underdoped data with a $T$%
-dependent BCS-like gap.\cite{gap}

A similar conclusion regarding the evolution of the gap with doping is
reached on the basis of the comparison of the calculated temperature
dependence of the relaxation time $\tau $ with the data. The divergence at $%
T_{c}$ shown in Figure 5 for samples near optimum doping together with the
temperature dependence of $\left| {\it \Delta }{\cal T}/{\cal T}\right| $
lead to the inescapable conclusion that there is a gap in YBa$_{2}$Cu$_{3}$O$%
_{7-\delta }$ with $(\delta \sim 0.1)$ which closes rapidly at $T_{c}$. On
the other hand, the notable {\em absence\ }of this divergence of $\tau $ for
underdoped YBa$_{2}$Cu$_{3}$O$_{7-\delta }$ $(\delta >0.15)$\cite{Demsar} is
consistent with the existence of a $T$-independent gap deduced from the
temperature dependence of $\left| {\it \Delta }{\cal T}/{\cal T}\right| $.

Plotting the temperature $T^{\ast }$ and $T_{c}$ as a function of $\delta $
in Figure 6, we find that $T^{\ast }$ drops with increasing doping in a
characteristic fashion, following the so-called ''pseudogap'' behaviour
reported by other experimental techniques\cite{Loram,NMR,Raman}. Near
optimum doping, the ''pseudogap'' becomes observable close to $T_{c}$. Such
a situation occurs in BCS\ superconductors where gap formation is a
collective effect and Cooper pairs form simultaneously with the formation of
a phase coherent collective state at $T_{c}$. However, from Figure 2c) and
Figure 3 there is some evidence of remaining QP\ excitations above $T_{c}$
even near $\delta \sim 0.1.$ The $T$-independent ''pseudogap'' thus appears
to be still present near optimum doping, and the BCS case is not quite
realized.

The underdoped state is distinctly different, with a $T$-independent normal
state pseudogap appearing at $T^{\ast }\gg T_{c}$. The changes in the
photoinduced transmission $\left| {\it \Delta }{\cal T}/{\cal T}\right| $ at 
$T^{\ast }$ in this case are simply a result of an increase population of
pairs with decreasing temperature. Since the observed changes in $\left| 
{\it \Delta }{\cal T}/{\cal T}\right| $ are associated only with pairing and
not phase coherence (which occurs at $T_{c}),$ we see no change in $\left| 
{\it \Delta }{\cal T}/{\cal T}\right| $ at $T_{c}$ in agreement with
predictions of pre-formed pairing models of the underdoped state\cite
{alex,emery}. The pseudogap in this case signifies the pair binding energy,
and its value is determined by the microscopic pairing mechanism. It is to
first approximation temperature independent and decreases with an inverse
law with increasing carrier concentration, which is consistent with
increased semi-classical (Debye-H\"{u}ckel) screening between charge
carriers \cite{alex}.

Since the temperature dependence of $\left| {\it \Delta }{\cal T}/{\cal T}%
\right| $ is actually exponential at high temperatures (Eq.4), it is more
appropriate to discuss the doping dependence of the magnitude of the energy
gap $\Delta $ rather than of $T^{\ast }.$ This avoids the somewhat arbitrary
criterion for determining $T^{\ast }$ of an asymptotic function. Such a plot
of $\Delta $ versus doping $\delta $ is shown in Figure 7. As predicted on
the basis of screening arguments above, ${\bf \Delta }$ appears to follow an
inverse law ${\bf \Delta }\propto 1/x$ where $x=0.6-\delta $ is proportional
to the carrier concentration.

The doping dependence of ${\bf \Delta }$ can be independently confirmed from
the same data set by plotting the low-temperature value of $\left| {\it %
\Delta }{\cal T}/{\cal T}\right| $ \ as a function of $\delta .$ As can be
seen from the formulae (4,6), the value of $\left| {\it \Delta }{\cal T}/%
{\cal T}\right| $ at $T=0$ is proportional to ${\cal E}_{I}/{\bf \Delta },$
from which the zero-temperature gap ${\bf \Delta }_{T=0}$ can be determined.
A scaled plot of $1/\left| {\it \Delta }{\cal T}/{\cal T}\right| $ $%
_{T\rightarrow 0}$ as a function of doping $\delta $ in Figure 7
independently confirms the inverse relation ${\bf \Delta }\propto
1/(0.6-\delta )$. We note that the observation of an inverse law is in
agreement with recent experiments on La$_{2-x}$Sr$_{x}$CuO$_{4}$\cite{Zhao}, 
\cite{Bi} suggesting that it may be a universal feature of the cuprates.

A somewhat surprising feature of the data is that they are described\ so
well by an isotropic gap over the whole range of doping. A $d$-wave gap is
apparently not consistent with either the temperature dependence of $\left| 
{\it \Delta }{\cal T}/{\cal T}\right| $ or the linear intensity dependence
of $\left| {\it \Delta }{\cal T}/{\cal T}\right| .$ It is also not
consistent with the observation of a single exponential decay of ${\it %
\Delta }{\cal T}/{\cal T}$\ with time, since in the $d$-wave case we would
expect a distribution of $\tau $s, and certainly no single divergence of $%
\tau $ at $T_{c}$ should be observed in the $d$-wave case. A possible
explanation for the absence of a $d$-wave signature is that in YBCO the gap
in the bulk of the sample is more or less isotropic in contrast to the
surface, where it may be more $d$-like\cite{Mueller}. Another possibility
for the apparent discrepancy between the present optical data and, for
example, Raman experiments\cite{Hackl} are the different weights that the
matrix elements have in the probing process. In the present experiments the
dipole transition matrix elements average over the whole Fermi surface (FS)
weighted by areas of large electron momentum. Anisotropy of the FS is not
emphasized in this case. This is different than in Raman measurements, where
the electronic scattering intensity is zero for an isotropic parabolic band,
and Raman intensity comes only from the anisotropic regions of the FS.

We conclude by noting that the comparison of the theoretically predicted
temperature dependence of the photoinduced absorption signal amplitude {\em %
and }relaxation time $\tau $ with the experimental results on underdoped and
optimally doped YBa$_{2}$Cu$_{3}$O$_{7-\delta }$ gives good {\em quantitative%
} agreement. The evolution of the gap from a temperature-independent $\Delta 
$ to a temperature-dependent collective gap $\Delta (T)$ is a particularly
striking feature of the data as is the apparent inverse dependence of the
gap magnitude with doping. Although the model calculation presented in this
paper was performed with the superconducting cuprates in mind, it is quite
general and can, for example, also be applied to a case of photoexcitation
experiments on CDW systems or narrow gap semiconductors. It could also be
extended for other non-equilibrium situations as might occur in
non-equilibrium superconducting high-$T_{c}$ devices for example.

\section{Acknowledgments}

We would like to acknowledge K.A.M\"{u}ller, R.Hackl and A.S. Alexandrov for
valuable comments and discussions. One of us (VVK) acknowledges support of
this work by RFBR Grant 97-02-16705 and by the Ministry of Science and
Technology of Slovenia.

\section{Appendix}

To estimate the electron-phonon relaxation rate we take into account the
electron-phonon collision integral\cite{Lifshitz}: 
\[
I_{ph-e}\{n_{\omega }\}=-4\pi \sum_{k}\left| M_{kk^{^{\prime }}}\right| ^{2}
\left[ f_{k}(1-f_{k^{^{\prime }}})n_{q}-f_{k^{^{\prime }}}(1-f_{k})\left(
n_{q}+1\right) \right] \delta (\epsilon _{k}-\epsilon _{k^{^{\prime
}}}+\omega _{q}) 
\]
$M_{kk^{^{\prime }}}$ is electron-phonon coupling constant and $n_{\omega }$%
, $f_{k}$ are phonon and electron distribution functions.

Note that we neglect coherence factors in the electron-phonon collision
integral. This is a reasonable approximation for the BCS case if the
characteristic energy of quasiparticles is large in comparison with the gap $%
T>{\bf \Delta }$. For underdoped case also, the normal state gap is
temperature independent and hence coherence factors are equal to 0. It is
easy to show that the collision integrals satisfy the conservation energy
law \cite{Allen}:

\[
2\sum_k\epsilon _k(\frac{\partial f_k} {\partial t})+\sum_q\omega _q(\frac{%
\partial n_\omega }{\partial t})=0. 
\]

To calculate electron-phonon collision integral we define ''electron-phonon
spectral function'' $\alpha ^{2}F$ : 
\[
\alpha ^{2}F(\epsilon ,\epsilon ^{^{\prime }},\Omega )=\frac{2}{N(0)}%
\sum_{k,k^{^{\prime }}}\left| M_{k,k^{^{\prime }}}\right| ^{2}\delta (\omega
_{q}-\Omega )\delta (\epsilon _{k}-\epsilon )\delta (\epsilon _{k^{^{\prime
}}}-\epsilon ^{^{\prime }}) 
\]

Following Allen\cite{Allen}, we suppose that $\alpha ^{2}F(\epsilon
,\epsilon ^{^{\prime }},\Omega )\simeq \alpha ^{2}F(\epsilon _{F},\epsilon
_{F},\Omega )=\alpha ^{2}F(\Omega )$. For the sake of simplicity we make the
Debye approximation for the electron-phonon spectral function $\alpha
^{2}F(\Omega )=\frac{\lambda \Omega ^{2}}{\Omega _{c}^{2}}$. 
\begin{eqnarray}
\sum_{q(\omega _{q}<2{\bf \Delta })}\omega _{q}I_{e-ph}\{n_{\omega }\}
&=&-2\pi N(0)\int_{0}^{2{\bf \Delta }}\Omega \alpha ^{2}F(\Omega )d\Omega
\int_{{\bf \Delta }}^{\infty }\{\left( f(\epsilon )-f(\epsilon +\Omega
)\right) n_{\Omega }-  \nonumber \\
&&f(\epsilon +\Omega )\left( 1-f(\epsilon )\right) \}d\epsilon  \eqnum{A1}
\end{eqnarray}

The integral Eq.(A1) can be easily evaluated: 
\begin{equation}
I_{3}\simeq \frac{4\pi k_{B}N(0)\lambda {\bf \Delta }^{4}}{\hbar ^{2}\Omega
_{c}^{2}}(T^{^{\prime }}-T)  \eqnum{A2}
\end{equation}

Formula (A2) leads to the Eq.(30) for electron-phonon relaxation rate.

\smallskip \newpage

Figure 1. The normalized photoinduced transmission ${\it \Delta }{\cal T}/%
{\cal T}$\ as a function of time delay for two different samples with a) $%
T_{c}=53$K and b) $T_{c}=90$K, each at three different temperatures. The
time-evolution of the traces was fitted (solid lines) using a model with a 
{\em single} exponential decay and a Gaussian temporal profile pump pulse.

Figure 2. a)\ The photoinduced transmission amplitude $\left| {\it \Delta }%
{\cal T}/{\cal T}\right| $ in YBa$_{2}$Cu$_{3}$O$_{7-\delta }$ as a function
of temperature for three different $\delta :$ for $\delta =0.48$ (open
squares)$,$ $\delta =0.18$ (solid triangles) and $\delta \simeq 0.1$ (open
circles). b) The solid line is a plot of the photoinduced transmission
amplitude as a function of temperature for a temperature-independent gap
using Expression (4) as a function of $T/T^{\ast }$. The data points are for
underdoped samples with $\delta =0.18$ with $T_{c}=77$ K (solid triangles), $%
\delta =0.32$ with $T_{c}=62$ K (open triangles), $\delta =0.44$ with $%
T_{c}=53$ K (solid circles) and $\delta =0.48$ with $T_{c}=48$ K (open
squares) respectively. c) The solid line is the calculated photoinduced
transmission amplitude from Eq. (6) for a temperature-dependent gap as a
function of normalized temperature $T/T_{c}$ and data for the near-optimally
doped samples $\delta \sim 0.1$ ($T_{c}=90$K) from Mihailovic et al.\cite
{Demsar} (open circles), Stevens et al. \cite{Stevens} (open squares) and
Han et al. \cite{Han} (solid triangles). Note: the original datapoints of
Ref.\cite{Han} were shifted by 7K in temperature. The solid lines in a) are
fits with Eq. (4) for $\delta =0.48$ and $\delta =0.18,$ and Eq. (6) for $%
\delta \simeq 0.1.$

Figure 3. The $T$-dependence of $\left| {\it \Delta }{\cal T}/{\cal T}%
\right| $\ for near optimally doped sample ($T_{c}$ = 89K) for three
different laser intensities. The data has been normalized with respect to
the energy density per unit volume deposited by the laser pulse, ${\cal E}%
_{I}$. The insert shows low temperature $\left| {\it \Delta }{\cal T}/{\cal T%
}\right| $\ vs. ${\cal E}_{I}$, where $\left| {\it \Delta }{\cal T}/{\cal T}%
\right| $ scales linearly with ${\cal E}_{I}$ and ${\cal E}_{0}=8\times
10^{-5}$ eV cell$^{-1}$spin$^{-1}.$

Figure 4. A plot of the temperature dependence of the photoinduced
transmission amplitude ( i.e. PE\ carrier density) for a pure $d$-wave gap
(Eq. (12)). The solid line is for a 3D case, while the dashed line is for a
2D case using the same parameters. (The ordinate scale is the same for both
curves.) The data (symbols) exhibit qualitatively different behaviour than
predicted by the $d$-wave model, irrespective of doping $\delta .$ The data
were scaled in the same way as in Fig.2. The open squares represent data for 
$\delta >0.15$ while the solid squares are for $\delta <0.1.$

Figure 5. The relaxation time $\tau $ as a function of temperature for a
temperature-dependent gap (Eq. (28)). The data are for the near-optimally
doped samples $\delta \sim 0.1$ ($T_{c}=90$K) from Han et al.\cite{Han}
(full circles) and Mihailovic et al.\cite{Demsar} (open circles). Note: the
data points of Han et al\cite{Han} were shifted by 7K in temperature.

Figure 6. $T_{c}$ (open squares) and $T^{\ast }$ (solid squares) as a
function of $\delta .$ $T^{\ast }$ is defined as the point where the
amplitude of\ signal $\left| {\it \Delta }{\cal T}/{\cal T}\right| $ drops
to 5\% of maximum of the $T$-dependent signal (the $T$-independent
contribution, which is assumed to be due to Allen relaxation is
subtracted).\ 

Figure 7. The dependence of the energy gap $\Delta $ on doping $\delta $.
The open squares correspond to values of $\Delta $ determined from fits to
the data using Eqs.(4) and (6), while the solid circles are obtained by
plotting the scaled magnitude of the inverse of the \ photoinduced
transmission at low temperature, $1/\left| {\it \Delta }{\cal T}/{\cal T}%
\right| _{T\rightarrow 0}$. In the latter case, the data were taken in a
single experimental run, with carefully controlled laser operating
conditions to ensure that $\Delta $ can be compared quantitatively. The
dashed line is a guide to the eye showing $\Delta \propto 1/x$ bahaviour.


\begin{references}
\bibitem{Han}  S.G.Han Z.V.Vardeny, K.S.Wong, O.G.Symco, G.Koren,
Phys.Rev.Lett. {\bf 65}, 2708 (1990)

\bibitem{Chwalek}  J.M.Chwalek, C.Uher, J.F.Whitaker, G.A.Morou and
J.A.Agostinelli, Appl.Phys.Lett. {\bf 58}, 980 (1991)

\bibitem{Albrecht}  W. Albrecht, Th.Kruse and H.Kurz, Phys.Rev.Lett. {\bf 69}%
, 1451 (1992)

\bibitem{Stevens}  C.J.Stevens, D.Smith, C.Chen, J.F.Ryan, B.Podobnik,
D.Mihailovic, G.A.Wagner and J.E.Evetts, Phys.Rev.Lett {\bf 78}, 2212 (1997).

\bibitem{Demsar}  D.Mihailovic, B.Podobnik, J.Demsar, G.Wagner and J.Evetts
(to be published in J.Phys.Chem Sol.,1998)

\bibitem{Mazin}  I.I.Mazin, A.I.Lichtenstein, O.Jepsen, O.K.Andersen and C.O
Rodriguez, Phys.Rev.B {\bf 49}, 9210 (1994)

\bibitem{comment}  I.I. Mazin, Phys. Rev. Lett. {\bf 80}, 3664 (1998) and
C.J.Stevens, D.Smith, C.Chen, J.F.Ryan, B.Podobnik, D.Mihailovic, G.A.Wagner
and J.E.Evetts, Phys. Rev. Lett. {\bf 80}, 3665 (1998)

\bibitem{Chekalin}  S.V. Chekalin et al, Phys. Rev. Lett., {\bf 67}, 3860,
(1991).

\bibitem{Brorson}  S.D.Brorson et al, Sol. Stat. Comm. {\bf 74}, 1305 (1990)

\bibitem{Allen}  P.B.Allen, Phys.Rev.Lett. {\bf 59}, 1460 (1987).

\bibitem{Basov}  D.N. Basov, R.Liang, B.Dabrowski, D.A. Bonn, W.N.Hardy,
T.Timusk, Phys. Rev. Lett., {\bf 77}, 4090, (1996).

\bibitem{Roth}  A. Rothwarth and B.N. Taylor, Phys. Rev. Lett., {\bf 19},
27, (1967).

\bibitem{Aronov}  A.G. Aronov and B.Z. Spivak, J. Low Temp. Phys., {\bf 29},
149, (1977).

\bibitem{Aronov1}  A.G. Aronov, M.A. Zelikman and B.Z. Spivak, Fiz. Tverd.
Tela, {\bf 18}, 2209, (1976).

\bibitem{approx}  Solving eq.(4) numerically without the approximate formula
for $n_{T}$, it can be shown that formula (6) is accurate within $<10\%$
over the whole temperature interval.

\bibitem{Schles}  Z. Schlesinger et al, Phys. Rev. Lett. {\bf 65}, 801,
(1990); L.D. Rotter et al, Phys. Rev. Lett. {\bf 67}, 2741, (1991).

\bibitem{Lifshitz}  J.M.Ziman ''{\it Electrons and Phonons}'' (Oxford
University Press, London 1960), E.M. Lifshits, L.P. Pitaevskii, ''{\it %
Fizicheskaia Kinetika}'' (Moscow, 1979).

\bibitem{Easley}  G.L.Easley, J.Heremans, M.S.Meyer, G.L.Doll and S.H.Liou,
Phys.Rev.Lett.{\bf 65, }3445 (1990).

\bibitem{Schmidt}  A. Schmidt and G. Schon, J. Low. Temp. Phys. {\bf 20},
207, (1975).

\bibitem{Tinkham}  M. Tinkham and J. Clarke, Phys. Rev. Lett. {\bf 28},
1366, (1972).

\bibitem{Mihailovic1}  D.Mihailovic, K.F.McCarty and D.S.Ginley, Phys.Rev.B 
{\bf 47}, 8910 (1993)

\bibitem{divergence}  The divergence in $\tau $ appears approximately 5-8 K
below $T_{c}$. Possibly this is due to a non-uniform temperature profile
within illuminated area.

\bibitem{Reitze}  D.H. Reitze, A.M. Weiner, A. Inam, and S. Etemand Phys.
Rev. B {\bf 46}, 14309, (1992).

\bibitem{gap}  Here the term {\em temperature-independent gap }means that
the gap does not vanish at $T_{c}$, but persists above $T_{c}$ It does not
exclude the possibility of a weakly temperature dependent gap.

\bibitem{Loram}  J.W. Loram et al, J. Superconductivity {\bf 7}, 243 (1994),
J. Loram et al, Phys. Rev. Lett {\bf 71}, 1740 (1993)

\bibitem{NMR}  G.V.M. Williams et al, Phys. Rev. B {\bf 51}, 16503 (1995), $%
{\sl ibid.}$ Phys. Rev. Lett. {\bf 78}, 721 (1997)

\bibitem{Raman}  G. Ruani and P. Ricci, Phys. Rev. B {\bf 55}, 93 (1997)

\bibitem{alex}  A.S. Alexandrov and N.F.Mott, ''{\it Polarons and Bipolarons}%
'' (World Scientific, 1995), A.S. Alexandrov and N.F. Mott, High Temperature
Superconductors and Other Superfluids (Taylor and Francis, London, 1994),
A.S.Alexandrov, V.V.Kabanov and N.F.Mott, Phys.Rev.Lett. {\bf 77}, 4796
(1996)

\bibitem{emery}  V.J.Emery, S.A.Kivelson, and O. Zahar, Phys. Rev.B {\bf 56}%
, 6120 (1997), V.J. Emery and S.A. Kivelson, Nature, {\bf 374}, 434, (1995).

\bibitem{Zhao}  K.A. Muller, G-M. Zhao, K. Conder, and H. Keller, J. Phys.:
Condens. Matter, {\bf 10}, L291 (1998).

\bibitem{Bi}  X.X.Bi and P.Ecklund, Phys.Rev.Lett. {\bf 70}, 2625 (1993).

\bibitem{Mueller}  K.A.M\"{u}ller, private communication

\bibitem{Hackl}  R. Nemetschek et al., Phys. Rev. Lett. {\bf 78}, 4834 (1997)
\end{references}
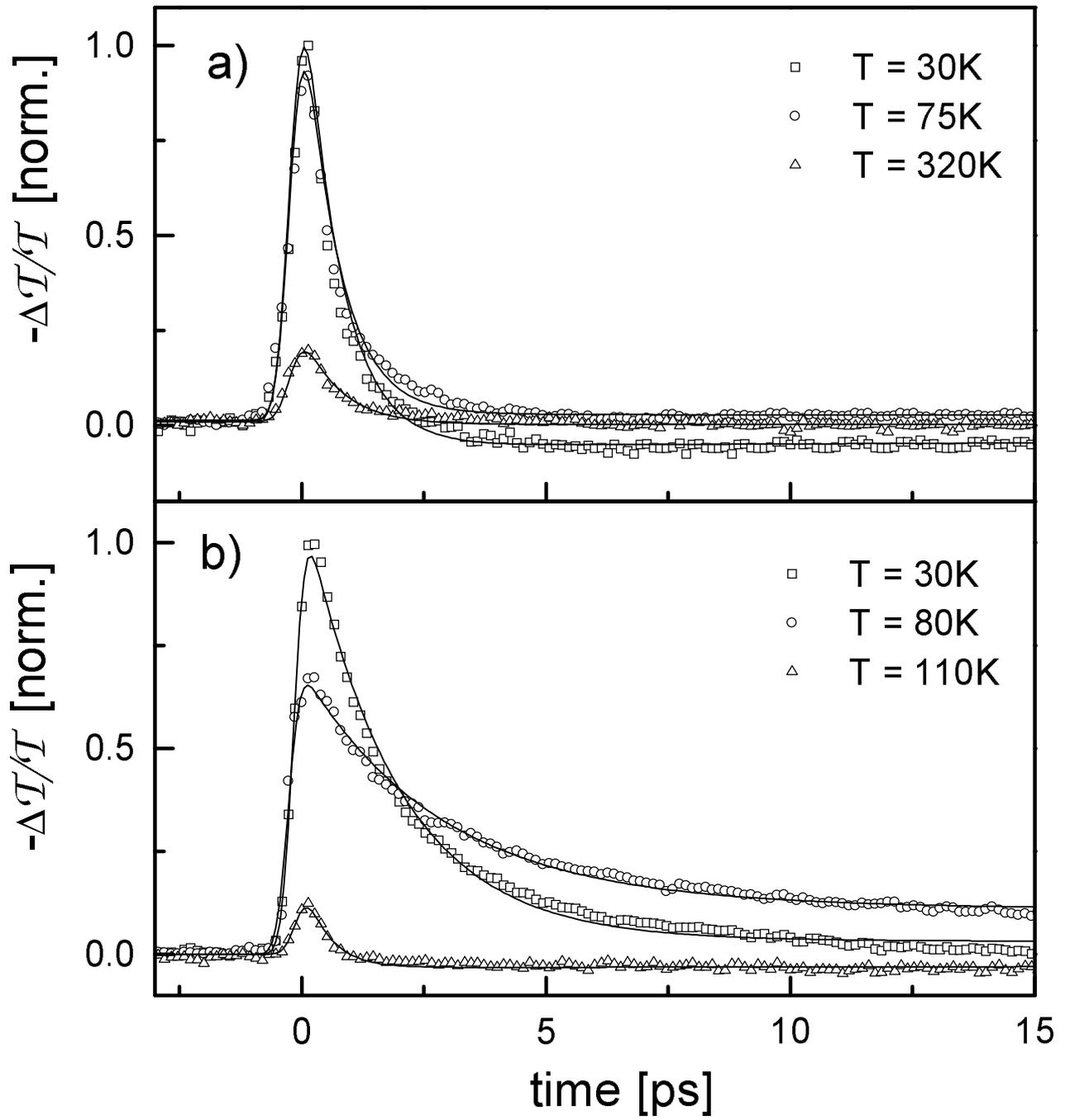
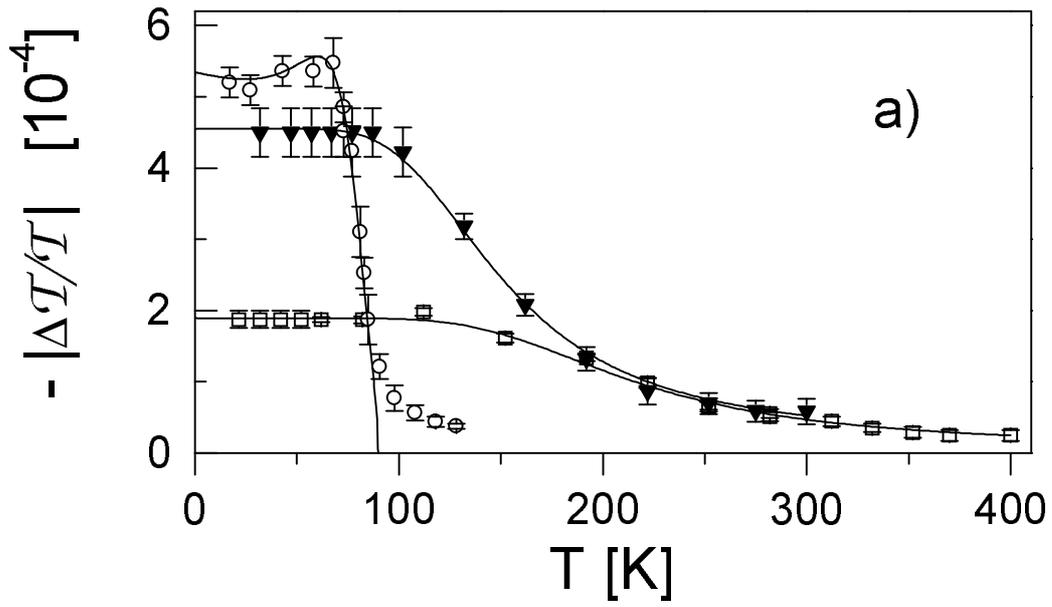
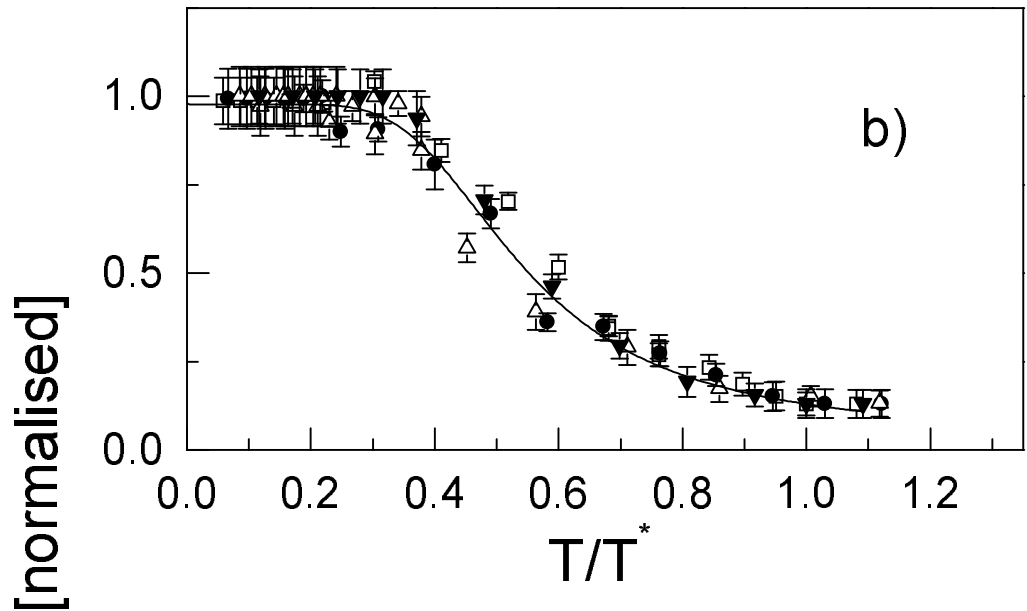
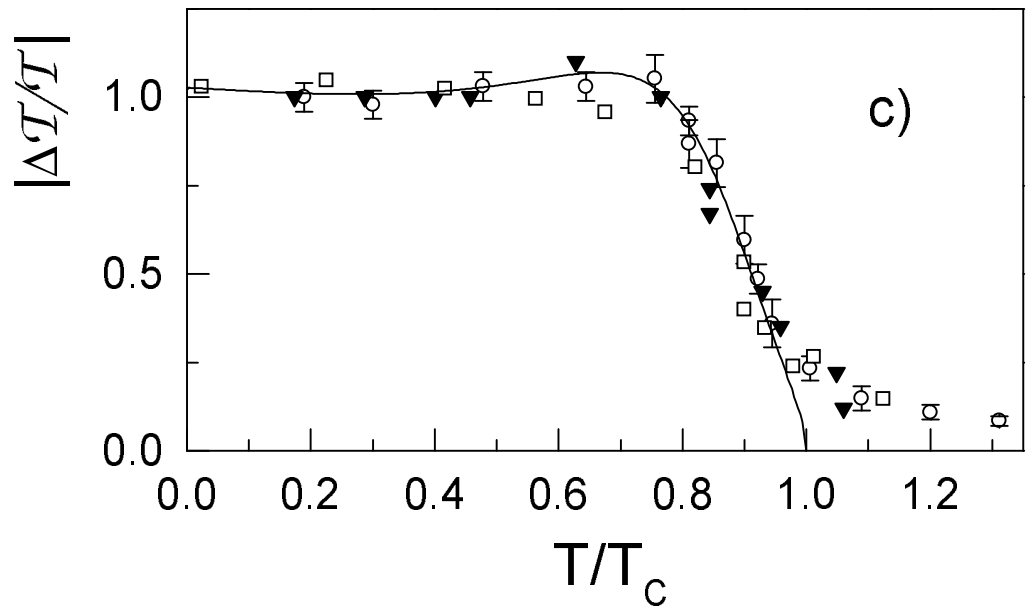
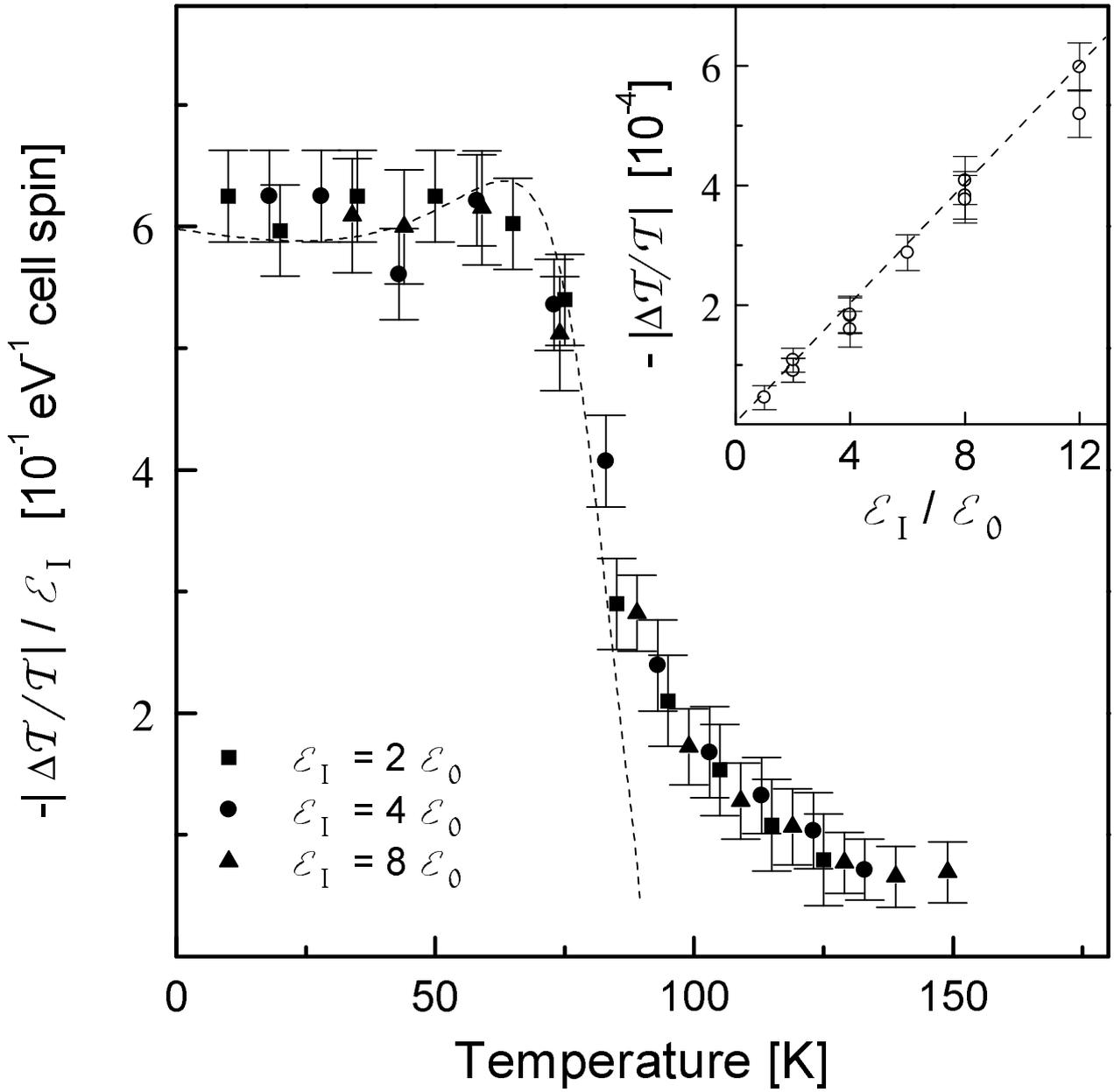
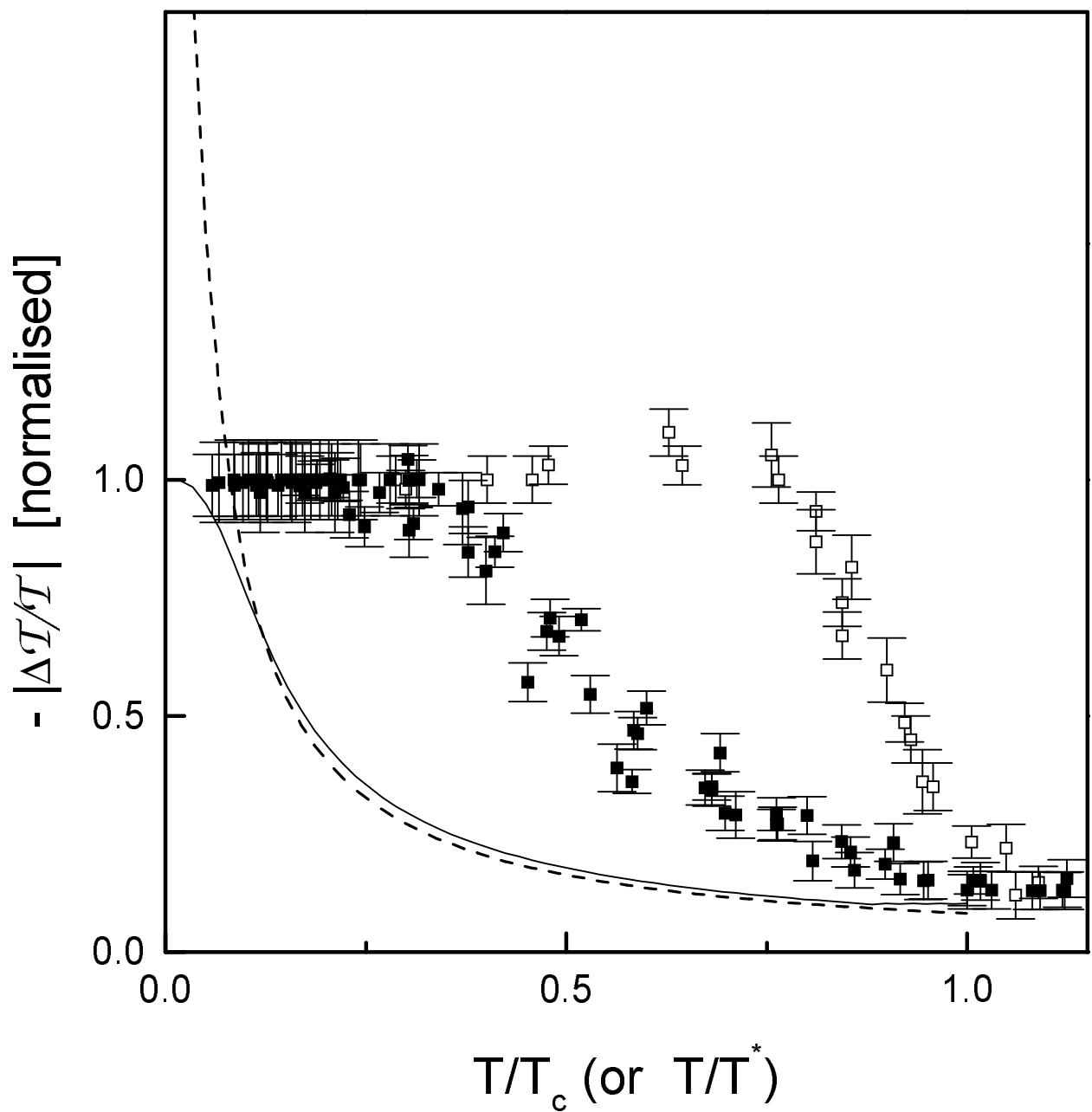
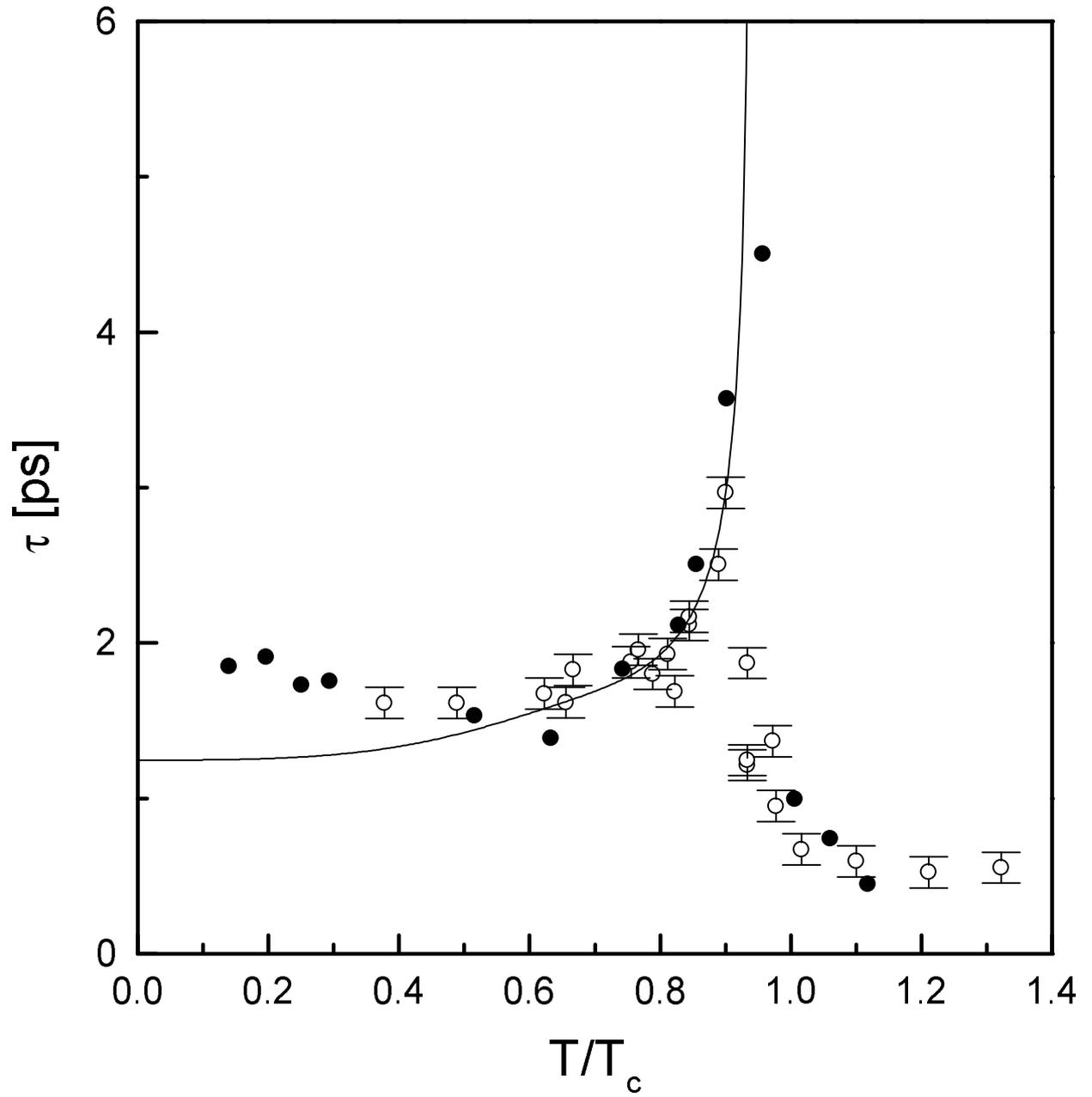
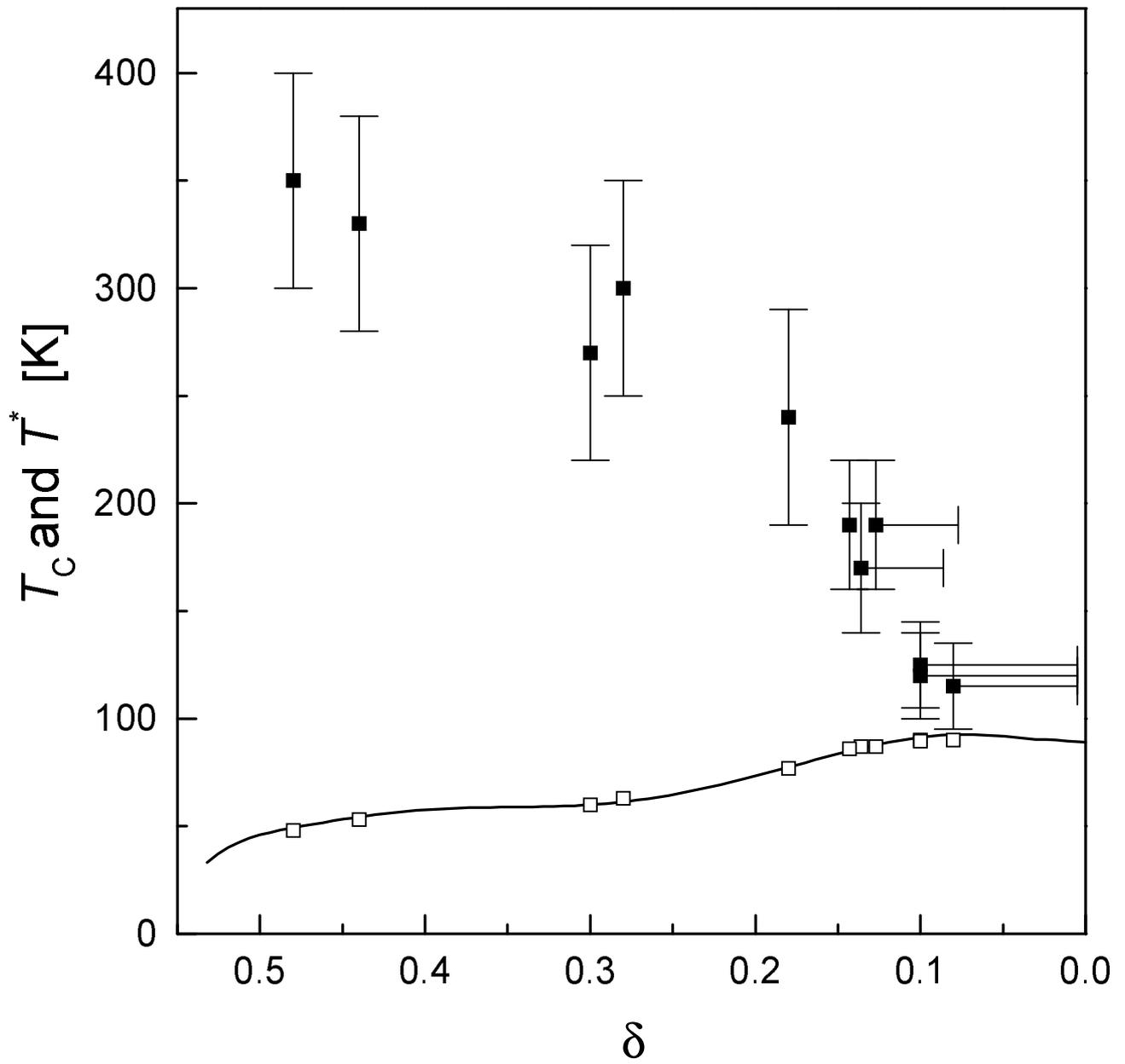
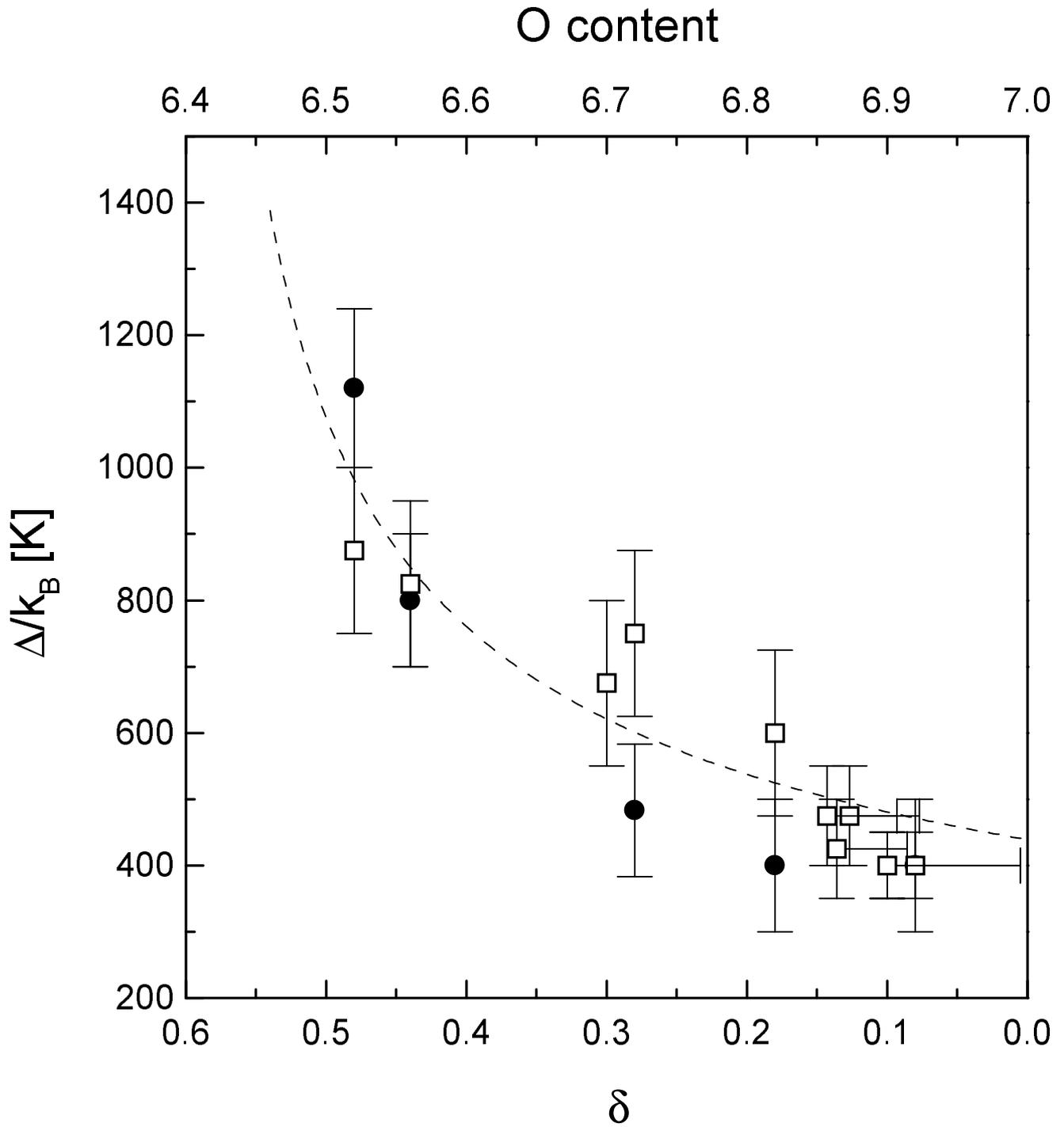
\end{document}